\documentclass[aps,prl,twocolumn,amsmath,amssymb,superscriptaddress,floatfix,showpacs]{revtex4}

\usepackage{graphicx}

\newcommand{\Q}{{\bf Q}}
\newcommand{\Sp}{{\bf S}}
\newcommand{\su}[1]{{$SU({#1})$}}

\begin{document}
\title{The quadrupolar phases of the  $S=1$ 
bilinear--biquadratic 
Heisenberg model on the triangular lattice}

\author{Andreas L\"auchli}
\affiliation{
Institut Romand de Recherche Num\'erique en Physique des Mat\'eriaux (IRRMA), CH-1015 Lausanne}

\author{Fr\'ed\'eric Mila}
\affiliation{ 
Institute of Theoretical Physics, Ecole Polytechnique F\'ed\'erale de Lausanne, CH-1015 Lausanne, Switzerland
}

\author{Karlo Penc}
\affiliation{
Research Institute  for  Theoretical Solid State  Physics   and
Optics, H-1525 Budapest, P.O.B.  49, Hungary}

\date{\today}

\begin{abstract}
Using mean-field theory, exact diagonalizations and \su{3} flavour theory, we have 
precisely mapped out the phase diagram of the $S=1$ bilinear--biquadratic 
Heisenberg model on the triangular lattice in a magnetic field, with emphasis on the 
quadrupolar phases and their excitations. In particular, we show that ferroquadrupolar 
order can coexist with short-range helical magnetic order, and that the antiferroquadrupolar 
phase is characterized by a remarkable 2/3 magnetization plateau, 
in which one site per triangle retains quadrupolar order while the other two are 
polarized along the field. 
Implications for actual $S=1$ magnets are discussed.
\end{abstract}


\maketitle

When discussing quantum magnets, it is useful to classify models according to whether or not the
ground state breaks the \su{2} symmetry. While simple examples of both cases are well known 
(long-range magnetic order for broken \su{2} symmetry, spin ladders for non-broken \su{2} symmetry),
a lot of activity is currently devoted to the problem of identifying more exotic ground states of
either type. In the context of non-broken \su{2} symmetry, the attention is currently focused
on the search for Resonating Valence Bond ground states in frustrated quantum magnets (for a review see \cite{claire}).

Regarding \su{2} broken ground states, the existence of
nematic order~\cite{spin_nematics} is well documented in a number of models, but its identification 
is faced with two difficulties: First of all, these models usually require 
four-spin exchange~\cite{ring} or biquadratic spin interactions \cite{harada} 
that are rather large for Mott insulators~\cite{FM}.
Besides, nematic order does not give rise to magnetic Bragg peaks, and there is a 
need for simple criteria that could help identifying nematic order experimentally.

In that respect, the recent investigation of NiGa$_2$S$_4$ \cite{nigas}, and the subsequent proposal by 
Tsunetsugu and Arikawa \cite{tsune} that some kind of antiferroquadrupolar (AFQ) order might be at the 
origin of the anomalous properties of that system, are very stimulating. 
They have investigated the AFQ phase of the $S=1$ Heisenberg model with bilinear and biquadratic exchange
on the triangular lattice, defined by the Hamiltonian:
\begin{equation}
  \mathcal{H} = J \sum_{\langle i,j \rangle} 
  \left[ 
  \cos \vartheta \ {\bf S}_i \cdot {\bf S}_j 
  + \sin \vartheta   \left({\bf S}_i \cdot {\bf S}_j\right)^2  \right] - h \sum_i S_i^z \;.
  \label{eq:ham}
\end{equation}
Using a bosonic description of the excitations,
they have investigated the zero-field case and shown in particular that: 
1) The magnetic structure factor has a maximum but no Bragg peak; 
2) The susceptibility does not vanish at zero temperature; 
3) The specific heat has the characteristic $T^2$ behaviour of 2D systems with broken \su{2} symmetry.
These features agree qualitatively with the properties of NiGa$_2$S$_4$, but several points remain
to be addressed. In particular, unlike the AFQ phase of the Heisenberg model studied 
in Ref.~\onlinecite{tsune}, the short-range magnetic fluctuations are incommensurate in NiGa$_2$S$_4$.
So the actual magnetic model describing NiGa$_2$S$_4$ remains to be worked out. 
Maybe more importantly, a more direct identification of quadrupolar order as being at the
origin of its properties would be welcome.

In the present paper, we address these points in the context of a thorough investigation of the $S=1$ 
bilinear-biquadratic Heisenberg 
model on the triangular lattice, Eq.~({\ref{eq:ham}).
We put special emphasis on the quadrupolar phases and on the effect of a magnetic field.

\paragraph{Quadrupolar operators and states:} To discuss quadrupolar (QP) order, it is useful to introduce the QP operator $\Q_i $
of components 
$(S_i^x)^2 - (S_i^y)^2$,
$(2(S_i^z)^2 -(S_i^x)^2-(S_i^y)^2)/\sqrt{3}$,
$S_i^x S_i^y + S_i^y S_i^x$,
$S_i^y S_i^z + S_i^z S_i^y$, and
$S_i^x S_i^z + S_i^z S_i^x$. 
In the finite Hilbert space of $S=1$, the biquadratic term can also be expressed by quadrupolar operators, 
and the Hamiltonian can be rewritten as
\begin{equation}
 \mathcal{H} = \sum_{\langle i,j \rangle} 
 \left[ \left(J_1-\frac{J_2}{2}\right) {\bf S}_i\cdot {\bf S}_j
 + \frac{J_2}{2} {\bf Q}_i \cdot {\bf Q}_j 
  +  \frac{4}{3} J_2  \right] 
\end{equation}
with $J_1 = J \cos \vartheta$ and $J_2 = J \sin \vartheta$.
Since $\Q_i\cdot\Q_j + \Sp_i \cdot \Sp_j=  -2 \mathcal{P}_{i,j} -2/3$,
and since the permutation operator $\mathcal{P}_{i,j}$ has \su{3} symmetry for spin 1, 
the Hamiltonian
is \su{3} symmetric for $J_1=J_2$ ($\vartheta=\pi/4$ and $-3\pi/4$).
It turns out to be convenient to choose the following time--reversal invariant basis of the \su{3} fundamental representation:
\begin{equation}
  |x\rangle=\frac{i |1\rangle- i|\bar 1 \rangle}{\sqrt{2}};\quad
  |y\rangle=\frac{|1\rangle+|\bar 1 \rangle}{\sqrt{2}};\quad
  |z\rangle= -i |0\rangle \,.
  \label{eq:xyzbasis}
\end{equation}
Quadupolar spin-states are then defined as  linear combinations with real amplitudes $d_\nu$ (such that $|{\bf d}|=1$)
\begin{equation}
 |Q({\bf d})\rangle 
 = d_x |x\rangle + d_y |y\rangle+ d_z |z\rangle 
 \label{eq:Qstate}
\end{equation}
$|Q({\bf d})\rangle$ is time--reversal invariant, which implies that 
$\langle Q({\bf d}) | {\bf S} |Q({\bf d})\rangle =0$, and it is a zero eigenvalue eigenstate of the operator
$\left({\bf d} \cdot  {\bf S}\right)^2$. It describes a state where the spin fluctuates mostly in the 
directions perpendicular to the vector ${\bf d}$, referred to as the {\it director}, which 
nicely illustrates the very heart of a spin 
nematic state: it has no magnetic moment, but nevertheless breaks \su{2} symmetry due to 
the presence of anisotropic spin 
fluctuations.

\begin{figure}[b]
  \begin{center}
  \includegraphics[width=3.5truecm]{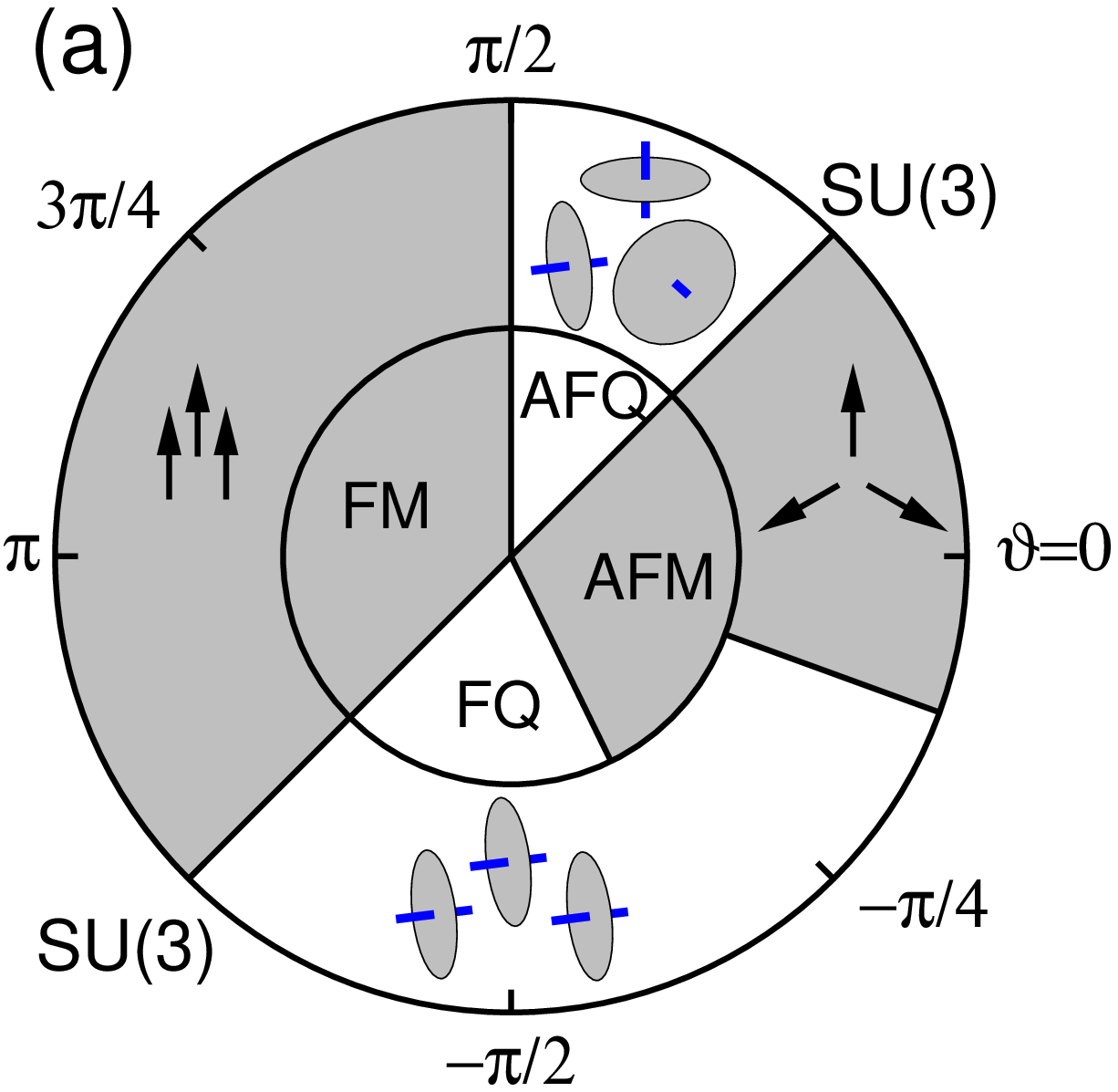}
  \includegraphics[width=4truecm]{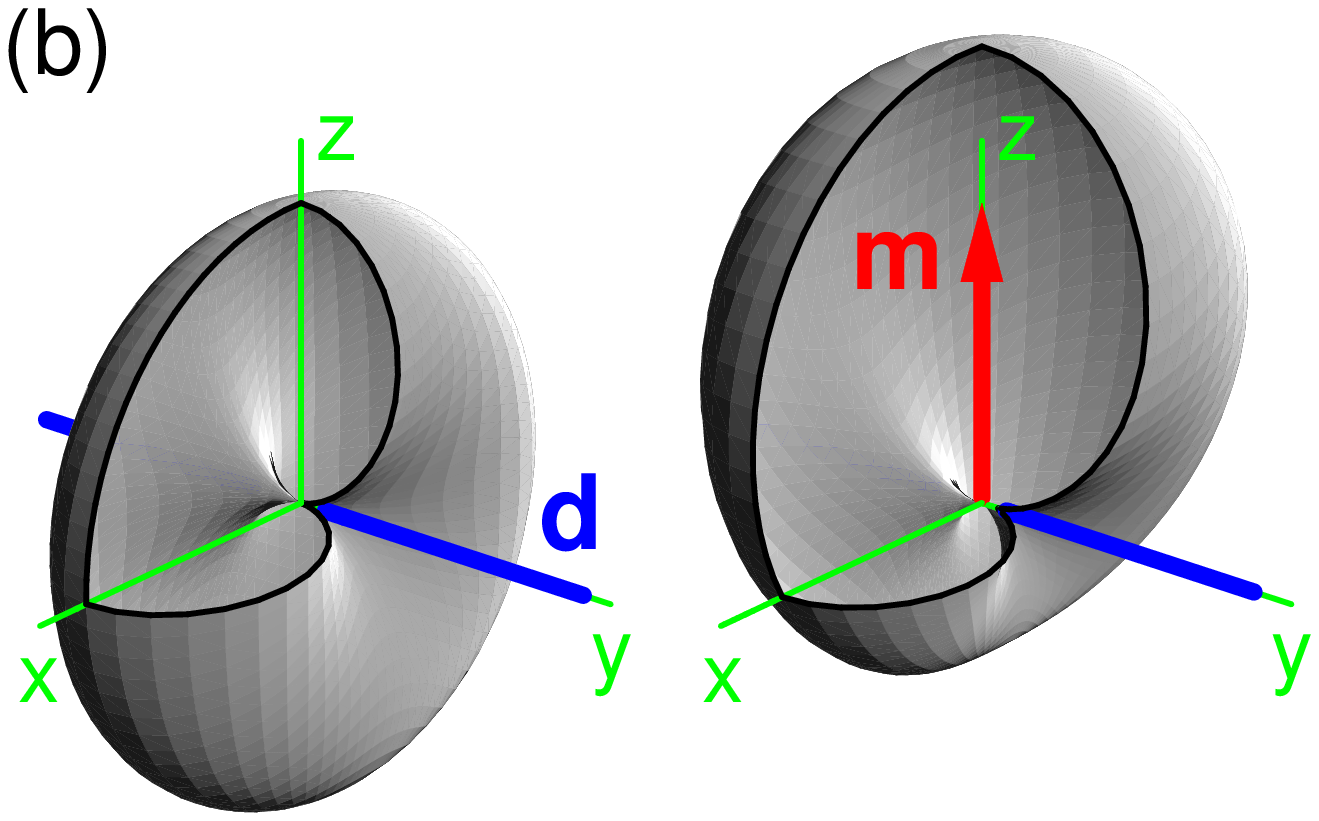}
  \end{center}
\caption{(color online) (a) Zero-field phase diagram. The inner circle is the variational result, 
the outer circle the exact--diagonalization one. 
The magnetic phases are shaded in gray.
(b) Probabilities of spin fluctuations $|\langle S(\hat{\bf n})|\psi\rangle|^2$ in the pure state
$\psi=|y\rangle$ (left) and in a state with finite magnetization (right). $| S(\hat{\bf n})\rangle$ is the
 coherent spin state pointing in direction $\hat{\bf n}$.
\label{fig:pd}}
\end{figure}

\paragraph{Zero-field phase diagram:} First, we construct the variational (mean-field) phase diagram in the
variational sub-space of site--factorized wave functions of the form  $\prod_j |Q({\bf d}_j)\rangle$, allowing complex ${\bf d}_j$'s 
\cite{ivanov03}, and assuming 3--sublattice long--range order.
Without magnetic field, we get four phases (Fig.~\ref{fig:pd}). Adjacent to the usual ferromagnetic (FM) phase, which is stabilized 
for $\pi/2<\vartheta<5\pi/4$, we find two QP phases. The expectation value of  $\Q_i \cdot \Q_j$ in the site--factorized wave function subspace is 
 \begin{equation}
  \langle \Q_i \cdot \Q_j \rangle = 2 |{\bf d}_i \cdot {\bf d}_j|^2 -
  2/3
   ; \quad i\neq j \;.
  \label{eq:QQdd}
 \end{equation} 
Since it induces a negative QP exchange, a negative biquadratic exchange tends to drive the directors collinear, 
leading to a stabilization of the ferroquadrupolar (FQ) state 
for $-3\pi/4<\vartheta<\Theta^{MF}_c$ with $\Theta^{MF}_c=\arctan(-2)\approx -0.35 \pi$. 
On the other hand, a positive biquadratic term induces a positive QP exchange, which is minimized with mutually perpendicular directors. On the triangular lattice, this is not frustrating since all bonds can be satisfied simultaneously by adopting a 3-sublattice configuration
with e.g. directors pointing in the $x$, $y$ and $z$ directions respectively,
a phase that can be called antiferroquadrupolar (AFQ). This is realized between the \su{3} point and the 
FM phase ($\pi/4<\vartheta<\pi/2$). 

For $\Theta^{MF}_c<\vartheta<\pi/4$, 
 we get the standard 3--sublattice 
120$^\circ$--antiferromagnetic (AFM) phase, but with a peculiarity: the spin length depends on $J_2/J_1$.
It is maximal ($|\langle {\bf S} \rangle|=1$) for $J_2=0$ and vanishes continuously 
as $|\langle S \rangle|\propto \sqrt{2J_1-|J_2|}$ at the FQ boundary, 
where the trial wave function becomes a QP state with the director 
perpendicular to the plane of  the spins. 
 Approaching} the \su{3} point, $|\langle {\bf S} \rangle|\rightarrow\sqrt{8}/3$, and the 
wave functions on the three sublattices become orthogonal. Actually, at this highly symmetric point 
any orthogonal 
set of wave functions is a good variational ground state. It includes the AFQ state as well, which is connected to 
the AFM state by a global \su{3} rotation.

\begin{figure}[b]
  \centering
  \includegraphics[width=8truecm]{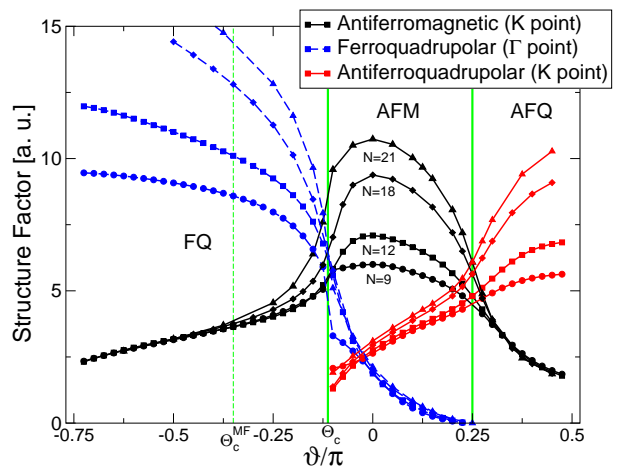}
\caption{(color online) Spin and quadrupolar structure factors at the $\Gamma$ and K points for different finite clusters.
The system sizes are labeled by the symbol type. 
\label{fig:sqED}}
\end{figure}
Next, we have performed finite--size exact diagonalization calculations on samples with up to 21 sites. 
In Fig.~\ref{fig:sqED}, we show the size dependence of the correlation functions 
associated with the FQ, AFM and AFQ order. More specifically we determine the structure factors 
$
\sum_j \exp[i \mathbf{k} \cdot 
\mathbf{r}_j] \langle \mathcal{C}_0\cdot \mathcal{C}_j\rangle$,
where $\mathcal{C}_j$ stands for the spin or quadrupolar operator at site $j$ 
and $\mathbf{k}$ is the $\Gamma$ or $K$ point in the Brillouin--zone
for the ferro or antiferro phases, respectively.
As can be clearly seen, the \su{3} point separates the AFM and AFQ phases, and the
$\vartheta$ dependence of the structure factors in the AFQ range is reminiscent
of that reported in the 1D model~\cite{andreas_1D}.
The phase boundary between the FQ and AFM phases is, on the other hand, strongly renormalized from the mean-field 
value $\Theta^{MF}_c\approx -0.35\pi$ to about $\Theta_c\approx -0.11\pi$ ($J_2\approx -0.4J_1$)~\cite{note_phaseboundary}.
We have also verified the presence of the appropriate 
Anderson towers of states
in the energy spectrum for the FQ, AFM, \su{3} AF, and AFQ phase \cite{longpaper}. Let us emphasize 
that we found no indication of disordered or liquid phases in that model.

\begin{figure}[t]
  \centering
  \includegraphics[width=7.5 truecm]{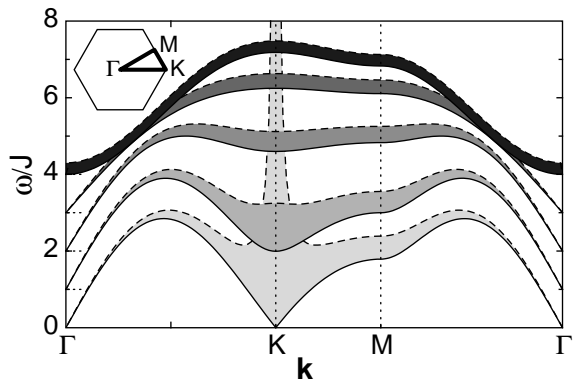}
\caption{Spin correlation function $S({\bf k},\omega)$ from flavor-wave calculation for $\vartheta=-3\pi/4$, $-5\pi/8$, $-\pi/2$, $-3\pi/8$, and $\Theta^{MF}_c$ from top to bottom (shifted by J) in the FQ phase. The dispersion $\omega(q)$ is the solid curve, the matrix element is the dashed line measured from the solid line. 
Left upper corner: the Brillouin zone.  \label{fig:sqomega}}
\end{figure}

\paragraph{Quadrupole waves:}
Since the usual spin--wave theory is not adequate to describe the excitations of QP phases, we 
use the flavour-wave theory of \cite{papanicolaou84}.
We associate 3  Schwinger--bosons $a_\nu$ to the 
states of  Eq.~(\ref{eq:xyzbasis}) and enlarge the fundamental representation on a site to a fully 
symmetric \su{3} Young-diagram consisting of an $M$ box long row. 
The spin operators are expressed as  $S^\alpha(j)= -i \epsilon_{\alpha\beta\gamma} a_\beta^{\dagger}(j) a_\gamma^{\phantom{\dagger}}(j)$. 
Condensing the bosons associated with the ordering then leads to a Holstein-Primakoff transformation. 
This approach is equivalent to the bosonic description of the AFQ phase in \cite{tsune},
so we will concentrate on FQ phase. To 
describe a state with all directors pointing in the $y$ direction, we let the 
$y$ bosons condense and
replace 
$a_y^{\dagger}$ and $a_y^{\phantom{\dagger}}$ by 
$(M-a_x^{\dagger} a_x^{\phantom{\dagger}}-a_z^{\dagger} a_z^{\phantom{\dagger}})^{1/2} $.
A $1/M$ expansion up to quadratic order in the Holstein-Primakoff bosons $a_x$ and $a_z$ 
followed by a standard
Bogoliubov transformation leads to:
\begin{equation}
  \mathcal{H} = \sum_{\nu=x,z}\sum_{\bf k} \omega_{\nu}({\bf k})
  \left[ \alpha_\nu^{\dagger}({\bf k}) \alpha_\nu({\bf k}) +
  1/2
\right]\,,
\end{equation} 
with $\omega_\nu({\bf k})= \sqrt{A^2_\nu({\bf k}) - B^2_\nu({\bf k})}$, 
$A_{\nu}({\bf k}) = 3 (J_1 \gamma({\bf k})- J_2)$,
$B_{\nu}({\bf k}) = 3 (J_2-J_1) \gamma({\bf k})$,
and $\gamma({\bf k})=\sum_{{\bf r}} \exp(i {\bf r}\cdot {\bf k})/6$,
where ${\bf r}$ spans the 6 neighbours of a site.
We get two branches of quadrupole waves associated to $a_x$ and $a_z$. In zero field, they
are degenerate throughout the entire Brillouin zone.

The imaginary part of the spin-spin 
correlation function 
(shown in Fig.~\ref{fig:sqomega}) 
is given by:
\begin{equation}
 S^{xx}({\bf k},\omega) =
  \frac{A_z({\bf k})+B_z({\bf k})}{\omega_z({\bf k})}
  \delta(\omega-\omega_z({\bf k})) \;.
\end{equation}
The $\nu=z$ branch contributes to 
$S^{xx}({\bf k},\omega) $ -- the spin fluctuations are perpendicular to both the 
director $y$ and the direction of the QP excitation $z$.
Correspondingly, the $\nu=x$ branch contributes to 
$S^{zz}({\bf k},\omega) $, which is equal to $S^{xx}({\bf k},\omega) $. $S^{yy}({\bf k},\omega)=0$ (it appears in higher order in $1/M$). 
Close to the $\Gamma$ point, both the dispersion of the flavor-wave and the correlation function are linear in $k$: 
$ \omega_\nu({\bf k}) = v  k$ and
 $ S^{xx}({\bf k})=S^{zz}({\bf k})=\chi v k 
$, where $\chi=1/[6(J_1-J_2)]$ is the mean--field susceptibility and $v=\sqrt{9J_2(J_2-J_1)/2}$.
As we approach the boundary to the AFM phase, $\omega_\nu({\bf k})$ softens and the spin structure factor diverges at the $K$ point of the Brillouin zone as
$ S({\bf k}) 
\propto \xi (1+\xi^2 |{\bf k}-{\bf k}_K|^2 )^{-1/2}$,
where the correlation length $\xi=1/\sqrt{10(\Theta_c^{MF}-\vartheta)}$ is associated with the short range easy--plane 120$^\circ$ AFM order. So FQ order can coexist with non-ferromagnetic short-range correlations,
a result to keep in mind when comparing with experiments.
 
\begin{figure}[bt]
  \centering
  \includegraphics[width=7.5truecm]{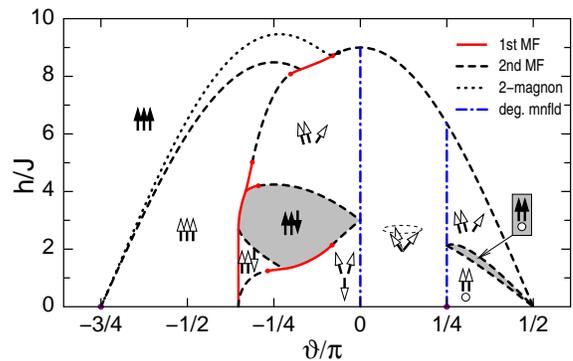}
\caption{(color online) Magnetic phase diagram. Solid (dashed) lines denote 1st  (2nd) order phase boundaries in the variational approach. The dotted line shows the exact boundary of the FQ phase (two magnon bound state formation, see Ref.\protect{\cite{longpaper}}). Along the dashed--dotted lines the variational solution is highly degenerate. The plateaux are shaded in gray. Filled arrows represent fully polarized magnetic moments, empty arrows partially polarized ones.
\label{fig:mpd}}
\end{figure}

\paragraph{Finite magnetic field:} 
In this case, the
variational phase diagram is surprisingly rich, as shown in Fig.~\ref{fig:mpd}. 

For $0<\vartheta<\pi/4$, the 3-sublattice AFM order gives rise to the chiral umbrella configuration up to full polarization. However, for negative $J_2$, a $m=1/3$ plateau occurs with up-up-down configuration, reminiscent of the plateau reported for the $S=1/2$ case \cite{Honecker}.
Interestingly enough, this magnetic configuration is also realized below the plateau close to the FQ phase. 
Above the plateau, the spin configuration is coplanar (see Fig.~\ref{fig:mpd}).

In the FQ phase, the directors turn perpendicular to the magnetic field \cite{ivanov03}, and the QP state is given 
by:
\begin{equation}
 |\psi_j\rangle= \cos\frac{\mu}{2} |Q(d_x,d_y,0)\rangle + i \sin\frac{\mu}{2} |Q(-d_y,d_x,0)\rangle \;.
 \label{eq:FQh}
\end{equation}
It develops a magnetic moment $m=\langle S^z\rangle =  \sin \mu$ parallel to the magnetic field by shifting the center of the fluctuations 
[c.f. Fig.~\ref{fig:pd}(b)] .
 The magnetization grows linearly with the magnetic field 
as $m=h/[6(J_1-J_2)]$. 
One of the two degenerate gapless modes acquires a gap proportional to 
the field, 
while the other one remains gapless -- it is the Goldstone mode 
associated with the 
rotation of the director in the $xy$ plane \cite{longpaper}.

\begin{figure}[b]
  \centering
  \includegraphics[width=8truecm]{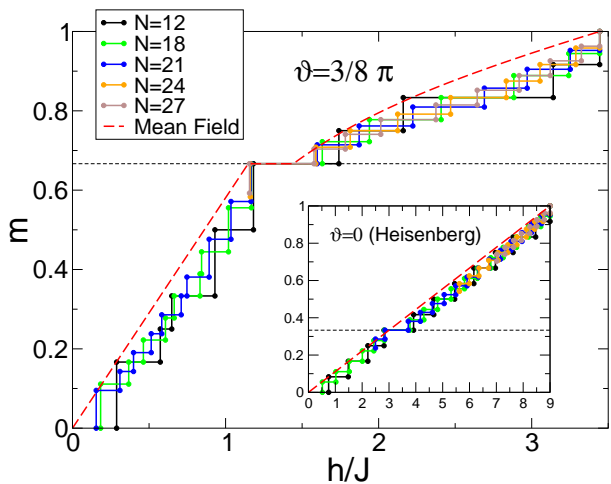}
  \caption{(color online) Magnetization curves obtained by exact diagonalization for $\vartheta=3\pi/8$, where the
    presence of a plateau at $m=2/3$ is confirmed. The inset shows a magnetization curve at
    the Heisenberg point with a small plateau stabilized at $m=1/3$. 
  \label{fig:magnetization}}
\end{figure}

The most surprising feature of the phase diagram is the magnetization plateau at $m=2/3$ that occurs starting from the AFQ phase ($\pi/4<\vartheta<\pi/2$) \cite{tsunem2o3}. This plateau is of mixed character: It is a $|110\rangle$  state, where
$|1\rangle$ is magnetic and $|0\rangle$ QP.  
It is stable between 
$h=3J_1$ and $3(4J_1-3J_2+\sqrt{9J_2^2-8J_1 J_2})/2$ according to the variational calculation, a result
confirmed by exact diagonalizations of the
magnetization for systems with up to 27 sites: There is indeed clear evidence of a plateau at 2/3, and the overall magnetization curve is in good agreement with the mean-field result, as shown in Fig.~\ref{fig:magnetization}.
Note that the occurrence of a plateau at 2/3 without a plateau at 1/3 is truly remarkable. Indeed, this is very
unlikely to occur for purely magnetic states since the 2/3 plateau would correspond to a higher commensurability
(5 up, 1 down in the simplest scenario) than the 1/3 one. This plateau can be considered as a characteristic 
of AFQ order.
Below the $m=2/3$ plateau, for $\pi/4<\vartheta<\pi/2$ and $h<3\cos\vartheta$ the variational ground state is the
deformed QP state of Eq.~(\ref{eq:FQh}), with two 
perpendicular
directors in the $xy$ plane, while on the third sublattice the QP state is $|0\rangle$. 
The magnetization is given by $m=2h/(9J_1)$.

\paragraph{Discussion:}
Let us put these results in experimental perspective. The occurrence of FQ order for negative biquadratic
exchange makes it a more likely candidate {\it a priori}. Indeed, a mechanism based on 
orbital quasi-degeneracy
has been shown to naturally lead to a {\it negative} biquadratic coupling \cite{orbitalbiquadratic}. In contrast, 
mechanisms leading to a large positive biquadratic coupling remain to be found. Otherwise, the FQ and AFQ 
have a lot in common, in particular
gapless modes, maxima but no Bragg peaks in the magnetic structure factor, 
a $T^2$ specific heat and a linear magnetization
at low field. Let us emphasize that the location of the maxima depends primarily on the bilinear exchange (sign, topology, range,...) and is {\it not} directly related to that of the gapless modes, which depends on the type of
QP order, hence on the sign of the biquadratic exchange.
The main difference is the presence of a remarkable magnetization $m=2/3$ plateau in the AFQ
phase. In addition, the removal of one of the gapless modes of the FQ state in a field should be
visible in the low temperature specific heat. Regarding NiGa$_2$S$_4$, it will be interesting to include 
further neighbor bilinear interactions into the present model to see if a QP phase with dominant incommensurate fluctuations
can be stabilized. The properties of the corresponding phase are expected to depend strongly on the sign of
the biquadratic coupling, and the results of the present paper should help in identifying such a phase.

 \acknowledgements 
We are pleased to acknowledge helpful discussions 
with 
J. Dorier,
P. Fazekas, 
T. Momoi, 
S. Nakatsuji,
N. Shannon, 
H. Shiba,
P. Sindzingre, 
H. Tsunetsugu, 
K. Ueda.
We are grateful for the support of the Hungarian OTKA Grant Nos. T049607 and K62280, of the MaNEP, and of the
Swiss National Fund. We acknowledge the allocation of computing time on the machines of the CSCS (Manno) 
and the LRZ (Garching).


\begin{thebibliography}{99}
\bibitem{note_phaseboundary} While we cannot exclude on
the basis of the present results that long-wave length fluctuations stabilize
the AFM order below that critical value, the presence of a level
crossing at this value of $\vartheta$ in the 9 and 21-site clusters with no significant
size dependence gives further support to this identification of $
\Theta_c$.
\end{thebibliography}
\end{document}